\documentclass[reqno]{article}
\usepackage{alltt}
\usepackage{graphicx}
\title{Quantum dynamics and state-dependent affine gauge fields on CP(N-1)}
\author{Peter Leifer}
\date{Cathedra of Informatics, Crimea State Engineering and
Pedagogical University, \\
21 Sevastopolskaya st., 95015 Simferopol, Crimea, Ukraine; \\Hermon Laboratories, Ltd. \\
Binyamina, 30500 Israel }
\begin{document}
\maketitle
\begin{abstract}
Gauge fields frequently used as an independent
construction additional to so-called wave fields of matter. This artificial
separation is of course useful in some applications (like Berry's interactions between the ``heavy" and ``light" sub-systems) but it is
restrictive on the fundamental level of ``elementary" particles and entangled states. It is
shown that the linear superposition of action states and non-linear dynamics
of the local dynamical variables form an oscillons of energy
representing non-local particles - ``lumps" arising together with their ``affine gauge potential" agrees with Fubini-Study metric.

I use the conservation laws of local dynamical
variables (LDV's) during affine parallel transport in complex projective
Hilbert space $CP(N-1)$ for twofold aim. Firstly, I formulate the variation
problem for the ``affine gauge potential" as system of partial differential
equations \cite{Le1}. Their solutions provide embedding quantum dynamics into
dynamical space-time whose state-dependent coordinates related to the qubit spinor
subjected to Lorentz transformations of ``quantum boosts" and ``quantum rotations".
Thereby, the problem of quantum measurement being reformulated as the comparison
of LDV's during their affine parallel transport in $CP(N-1)$, is inherently
connected with space-time emergences.
Secondly, the important application of these fields is the completeness of quantum theory. The EPR and Schr\"odinger's Cat paradoxes are discussed from the point of view of the restored Lorentz invariance due to the affine parallel transport of local Hamiltonian of the soliton-like field.
\end{abstract}


\section{Introduction}
Space-time geometry was established during four revolutionary
steps: Euclidean axiomatization of 3-geometry, summarizing previous
\emph{mechanical hard body} measurements, Galileo-Newton's dynamics postulating
absolute space and time structure, Einstein's discovery of
pseudo-Euclidean space-time 4-geometry which physically based on
measurements by means of \emph{classical electromagnetic field}, and
Einstein's discovery of pseudo-Riemannean space-time 4-geometry
relies upon specification of same kind of measurements in
gravitation field. One sees that the modeling the space-time
geometry is boosted up by the development of measuring process in
more general physical conditions.

Classical physics of system of material points frequently uses
multi-dimensional generalization of space-time (configuration space), but such generalization deems to be merely artificial construction whereas space-time is treated fundamentally. If one, however, takes into account that the notion of material point has
very limited sense in quantum theory then one will see just state space of quantum system has fundamental sense. It seems that the consistent development of quantum theory
requires a new geometry of quantum state space and its new connection with space-time.

Probably the biggest mistake of our scientific paradigm is the
assumption that we live in Universe coinciding with single 4D space-time
curved locally by a matter distribution. This illusion is strongly supported by our
every day experience and astronomy. Freely propagating quantum
particles and macroscopic bodies clearly show that our ``empty"
space is absolutely transparent and the pseudo-Euclidean or
pseudo-Riemannian space-time are good enough models for both macro-
and microscopic physical theories. Nevertheless, there are some
serious \textbf{logical} reasons to revise last assumption.

One of them is highly desirable agreement between quantum
behavior and classical relativity (both special and general) that
looks like the far distant future \cite{Penrose}. All known attempts
to reach the harmony (strings, super-gravity, e.g.) lead to unobservable predictions
(say, super-partners) and requirements entirely to change the basis of our scientific paradigm: instead of the observer independence one uses so-called the anthropic principle, instead of Universe was invoked so-called Multiverse concept. I propose much more modest approach.

I assume that Universe is represented by infinite dimension action state space and the space-time being macroscopically observable
as global pseudo-Riemannian manifold emerges due to projection during quantum measurement.
It means that quantum measurement may be used as operational procedure for ``marking" non-local quantum objects. I would like to recall that coordinatization
of classical events by means of {\it classical electromagnetic
field} is based on the distinguishability (separability), i.e. individualization of
pointwise material points. However we loss the possibility to distinguish (non-local) quantum objects by means of {\it quantum fields} in space-time and, hence it is impossible directly to identify them in space-time. It seems to be reasonable to assume that space-time has ``\textbf{granular structure} that respects Lorentz symmetry" only locally \cite{Bonder}.

\section{Space-time emergences due to the qubit-encoding of quantum measurement}
Generally, it is important to understand that the problem of identification of physical objects is the root problem even in classical physics and that its recognition gave to Einstein the key
to formalization of the relativistic kinematics and dynamics.
Indeed, only assuming the possibility to detect locally an approximate
coincidence of two pointwise events of a different nature it is
possible to build full kinematic scheme and the physical geometry of
space-time \cite{Einstein1,Einstein2}. As such the ``state" of the
local clock gives us local coordinates - the ``state" of the
incoming train. In the classical case the notions of the ``clock"
and the ``train" are intuitively clear and approximately may be identified with material points or even with space-time points. This supports the illusion that material bodies present in space-time (Einstein emphasized that it is not so!). Furthermore, Einstein especially notes that he did not discuss the inaccuracy of the
simultaneity of two {\it approximately coinciding events} that should
be overcame by some abstraction \cite{Einstein1}. This abstraction
is of course the neglect of finite sizes (and all internal degrees
of freedom) of the both real clock and train. It gives the
representation of these ``states" by mathematical points in
space-time. Thereby the local identification of two events is the
formal source of the classical relativistic theory. However quantum object requires especial embedding in space-time and its the identification with space-time point is impossible since the localization of quantum particles and generally even macroscopic space-time separation are state-dependent
\cite{Einstein2,W,Heg,A98}. Hence the identification of quantum objects requires a physically motivated operational procedure
with corresponding mathematical description.

Therefore the reason of deep conflict between general relativity
(GR) and quantum theory (QT) lies in space-time non-locality of
quantum systems (say, so-called entangled EPR pears), and as a
consequence, the problem of identification and comparison (measurement)  of
holistic quantum states by the intrinsic means of quantum theory. Besides
this there are the set of pure ``quantum paradoxes" (EPR, Zeno, ``interaction-free measurement'', etc., that require adequate (in fact - deterministic) description of non-local quantum objects. On the technical level \textbf{determinism} is equal to \textbf{locality} but in quantum area only functional, i.e. state space localization of quantum states is
possible \cite{Kryukov}. The fundamental question is as follows: \textbf{how to ``translate" the state space localization into ordinary space-time deterministic predictions} (science predictions have a price only in this case). I will show that technically this question may be formulated as follows: \textbf{what is the most natural manner of embedding quantum dynamics (in state space) into dynamics of the two-level detector}. My answer is based on the two postulates:

 1. ``Super-relativity" : \textbf{coset deformation of quantum state may be created and compensated by some physical fields} and

 2. ``Dynamical space-time structure emergence" :\textbf{quantum measurement of the local dynamical variables may be encoded by the local Lorentz transformations of the qubit spinor representing the state of two-level detector into local space-time}. Thereby local space-time structure accompanying quantum dynamics may be established.

I formulate the quantum measurement problem as a comparison of quantum LDV's (instead of
the comparison of quantum states) with help of their affine parallel transport
in $CP(N-1)$. Therefore one should deal with geometry
of the quantum state space, since we face with the measurement problem by
means of \emph{quantum fields}. I would like to emphasize: by means
of quantum fields rather in sense of de Broglie - Schr\"odinger than
by the means of \emph{secondary quantized fields} in the sense of
Dirac. We should remember that Schr\"odinger wrote his equation
initially just not in space-time but in the configuration space of a
system of material points \cite{Schr1}. However, material point is
good and very prolific abstract notion in classical theory but this
notion is terribly misleading in quantum area.

\section{Oscillons of the action states}
Each action state of quantum system is a quantum motion in some ``dynamical order"  defined, say, be some Lagrangian or action functional.
The state vector of this quantum motion thereby may be treated as ``order parameter" belonging to Hilbert space  - ``the space of the order parameter". This motion appears as excitations of global vacuum defined by the self-consistent cosmic potential $\Phi_U = c^2$ of Universe that may locally oscillate and create quantum
particles, solitons, unparticles, etc., under some conditions that should be especially established. Thus cosmic potential
$\Phi_U = c^2$ forms some global vacuum $|\Phi_U> = |\hbar 0>$ whose
perturbation by the action operator
\begin{eqnarray}
\hat{S}=\hbar A({\hat \eta^+} {\hat \eta}).
\end{eqnarray}
creates matter in some superposition (generalized coherent) state
\begin{eqnarray}
|F>=\sum_{a=0}^{\infty} f^a| \hbar a>,
\end{eqnarray}
where $|\hbar a>=(a!)^{-1/2} ({\hat \eta^+})^a|\hbar 0>$
constituting $SU(\infty)$ multiplete of the Planck's action quanta
operator $\hat{S_P}=\hbar {\hat \eta^+} {\hat \eta}$ with the
spectrum $S_a=\hbar a$ in the separable Hilbert space $\cal{H}$.
Formally these oscillations may be represented by the superposition
in infinite dimension manifold of the \emph{Planck's oscillators of
action}. Then well known relation of Einstein-de Broglie being
rewritten as follows $\frac{m}{\omega}=\frac{\hbar}{c^2}$ shows that
only fixed relations of mass to frequency are acceptable on the
fundamental level. Then the both mass $m$ and frequency $\omega$
should arise as intrinsically quantum values but not as in fact free
classical parameters. One may think about this model as some
abstract realization of the de Broglie ensemble of ``weights
suspended on springs'' serves as ``a crude analogue to a parcel of
energy" \cite{dB1}. In the de Broglie model the non-homogeneous
distribution of the weights on the disk was given ``by hands", but I
seek to find equations naturally describing similar distribution.

I assume that action states $|F>$ correspond to extremals of some least action problem and describe stationary quantum motion. These states do not
gravitate since they don't posses mass/energy. Therefore their linear superposition is robust and the rays of these generalized coherent states (GCS)
will be main building blocks of the model. Only velocities of variation of these states given by local dynamical variables correspond materialized particles, etc. For simplicity I will discuss here  $N$-dimension version of the model.

\section{Geometry of the quantum evolution and/or measurement}
Let me assume that ``ground
state" $|G>=\sum_{a=0}^{N-1} g^a|\hbar a>$ is a solution of some the least action problem.
Since any action state $|G>$ has isotropy group
$H=U(1)\times U(N)$ only the coset transformations $G/H=SU(N)/S[U(1)
\times U(N-1)]=CP(N-1)$ effectively act in $\cal{H}$. Therefore the
ray representation of $SU(N)$ in $C^N$, in particular, the embedding
of $H$ and $G/H$ in $G$, is a state-dependent parametrization.
Hence, there is a diffeomorphism between the space of the rays
marked by the local coordinates in the map
 $U_j:\{|G>,|g^j| \neq 0 \}, j>0$
\begin{equation}
\pi^i_{(j)}=\cases{\frac{g^i}{g^j},&if $ 1 \leq i < j$ \cr
\frac{g^{i+1}}{g^j}&if $j \leq i < N-1$}
\end{equation}\label{coor}
and the group manifold of the coset transformations
$G/H=SU(N)/S[U(1) \times U(N-1)]=CP(N-1)$. This diffeomorphism is
provided by the coefficient functions $\Phi^i_{\alpha}$ of the local
generators (see below).

The ``ground
state" $|G>=\sum_{a=0}^{N-1} g^a|\hbar a>$ may be expressed in local coordinates
as follows: for $a=0$ one has
\begin{eqnarray}\label{4}
g^0(\pi^1_{j(p)},...,\pi^{N-1}_{j(p)})=(1+
\sum_{s=1}^{N-1}|\pi^s_{j(p)}|^2)^{-1/2}
\end{eqnarray}
and for $a: 1\leq a = i \leq N-1$ one has
\begin{eqnarray}\label{5}
g^i(\pi^1_{j(p)},...,\pi^{N-1}_{j(p)})= \pi^i_{j(p)}(1+
\sum_{s=1}^{N-1}|\pi^s_{j(p)}|^2)^{-1/2}.
\end{eqnarray}
Then the velocity of the ground state evolution relative ``world
time" $\tau$ is given by the formula
\begin{eqnarray}\label{6}
|\Psi> \equiv |T> =\frac{d|G>}{d\tau}=\frac{\partial g^a}{\partial
\pi^i}\frac{d\pi^i}{d\tau}|\hbar a>+\frac{\partial g^a}{\partial
\pi^{*i}}\frac{d\pi^{*i}}{d\tau}|\hbar a> \cr
=|T_i>\frac{d\pi^i}{d\tau}+|T_{*i}>\frac{d\pi^{*i}}{d\tau}=H^i|T_i>+H^{*i}|T_{*i}>,
\end{eqnarray}
is the tangent vector to the evolution curve $\pi^i=\pi^i(\tau)$,
where
\begin{eqnarray}\label{42}
|T_i> = \frac{\partial g^a}{\partial \pi^i}|\hbar a>=T^a_i|\hbar a>,
\quad |T_{*i}> = \frac{\partial g^a}{\partial
\pi^{*i}}|\hbar a>=T^a_{*i}|\hbar a>.
\end{eqnarray}
Thereby state vector $|\Psi> \equiv |T>$ giving velocity of evolution,
is represented by the tangent vector to the projective Hilbert space $CP(N-1)$ where
instead of an arbitrary parameters $X= (x_1,...,x_p)$ of the
Hamiltonian I use intrinsic local projective coordinates
$\pi^k_{(j)}=\frac{g^k}{g^j}$ of the quantum states. Note, it is close (but not identical)
to the Berry's intuitive
analogy between the tangent vector $\vec{e}$ to the some sphere
$S^2$ and the state vector $|\phi (X)>=|\phi (x_1,...,x_p)>$ whose
time dependence is generated by the periodic Hamiltonian
$\hat{H}(X(t)):X(T)=X(0)$. The parallel
transport of $|\Psi>$ is required to be in agreement with the Fubini-Study metric
\begin{equation}\label{8}
G_{ik^*} = [(1+ \sum |\pi^s|^2) \delta_{ik}- \pi^{i^*} \pi^k](1+
\sum |\pi^s|^2)^{-2} \label{FS}.
\end{equation}
Then the affine connection
\begin{eqnarray}\label{9}
\Gamma^i_{mn} = \frac{1}{2}G^{ip^*} (\frac{\partial
G_{mp^*}}{\partial \pi^n} + \frac{\partial G_{p^*n}}{\partial
\pi^m}) = -  \frac{\delta^i_m \pi^{n^*} + \delta^i_n \pi^{m^*}}{1+
\sum |\pi^s|^2} \label{Gamma}
\end{eqnarray}
takes the place of the gauge potential of the non-Abelian type
playing the role of the covariant instant renormalization of the
dynamical variables during general transformations of the quantum
self-reference frame \cite{Le2}. Berry's parallel transport is not affine
and this acts in the fibre bundle of local reference frame rather in the
tangent bundle.

Velocity of the $|\Psi>$ variation is given by the equation
\begin{eqnarray}\label{43}
|A> &=&\frac{d|\Psi>}{d\tau} \cr &=&
(B_{ik}H^i\frac{d\pi^k}{d\tau}+B_{ik^*}H^i\frac{d\pi^{k*}}{d\tau}
+B_{i^*k}H^{i^*}\frac{d\pi^k}{d\tau} +B_{i^*
k^*}H^{i^*}\frac{d\pi^{k*}}{d\tau})|N>\cr &+&
(\frac{dH^s}{d\tau}+\Gamma_{ik}^s
H^i\frac{d\pi^k}{d\tau})|T_s>+(\frac{dH^{s*}}{d\tau}+\Gamma_{i^*k^*}^{s*}
H^{i*}\frac{d\pi^{k*}}{d\tau})|T_{s*}>,
\end{eqnarray}
where I introduce the matrix $\tilde{B}$ of the second quadratic
form whose components are defined by following equations
\begin{eqnarray}\label{45}
B_{ik}|N> =\frac{\partial |T_i>}{\partial \pi^k}-\Gamma_{ik}^s|T_s>,
\quad B_{ik^*}|N> = \frac{\partial |T_i>}{\partial \pi^{k*}} \cr
B_{i^*k}|N> =\frac{\partial |T_{i*}>}{\partial \pi^k}, \quad B_{i^*
k^*}|N> = \frac{\partial |T_{i*}>}{\partial
\pi^{k*}}-\Gamma_{i^*k^*}^{s*}|T_{s*}>
\end{eqnarray}
through the state $|N>$ normal to the ``hypersurface'' of the ground
states. Assuming that the ``acceleration'' $|A>$ is gotten by the
action of some linear Hamiltonian $\hat{H}$ describing the
evolution (say, during a measurement), one has the ``Schr\"odinger equation
of evolution"
\begin{eqnarray}\label{12}
\frac{d|\Psi>}{d\tau}&=&-i\hat{H}|\Psi> \cr
&=&(B_{ik}H^i\frac{d\pi^k}{d\tau}+B_{ik^*}H^i\frac{d\pi^{k*}}{d\tau}
+B_{i^*k}H^{i^*}\frac{d\pi^k}{d\tau} +B_{i^*
k^*}H^{i^*}\frac{d\pi^{k*}}{d\tau})|N> \cr &+&
(\frac{dH^s}{d\tau}+\Gamma_{ik}^s
H^i\frac{d\pi^k}{d\tau})|T_s>+(\frac{dH^{s*}}{d\tau}+\Gamma_{i^*k^*}^{s*}
H^{i*}\frac{d\pi^{k*}}{d\tau})|T_{s*}>.
\end{eqnarray}

I should emphasize that ``world time" is the  time of evolution from
the one GCS to another one which is physically distinguishable.
Thereby the unitary evolution of the action amplitudes generated by
(\ref{12}) leads in general to the non-unitary evolution of the tangent
vector to $CP(N-1)$ associated with ``state vector" $|\Psi>$ since
the Hamiltonian $\hat{H}$ is non-Hermitian and its expectation
values is as follows:
\begin{eqnarray}\label{13}
<N|\hat{H}|\Psi>&=&
i(B_{ik}H^i\frac{d\pi^k}{d\tau}+B_{ik^*}H^i\frac{d\pi^{k*}}{d\tau}
+B_{i^*k}H^{i^*}\frac{d\pi^k}{d\tau} +B_{i^*
k^*}H^{i^*}\frac{d\pi^{k*}}{d\tau}),\cr <\Psi|\hat{H}|\Psi>&=&
iG_{p^*s}(\frac{dH^s}{d\tau}+\Gamma_{ik}^s
H^i\frac{d\pi^k}{d\tau})H^{p*}+iG_{ps^*}(\frac{dH^{s*}}{d\tau}+\Gamma_{i^*
k^*}^{s*} H^{i^*}\frac{d\pi^{k*}}{d\tau})H^p\cr
&=&i<\Psi|\frac{d}{d\tau}|\Psi>.
\end{eqnarray}
The minimization of the $|A>$ under the transition from point $\tau$
to $\tau+d\tau$ may be achieved by the annihilation of the
tangential component
\begin{equation}
\frac{dH^s}{d\tau}+\Gamma_{ik}^s H^i\frac{d\pi^k}{d\tau}=0, \quad
\frac{dH^{s*}}{d\tau}+\Gamma_{i^* k^*}^{s*}
H^{i^*}\frac{d\pi^{k*}}{d\tau}=0
\end{equation}
i.e. under the condition of the affine parallel transport of the
Hamiltonian vector field. The last equations in (\ref{13}) shows that the
affine parallel transport of $H^i$ agrees with Fubini-Study metric
(\ref{8})  leads to Berry's ``parallel transport" of $|\Psi>$.
Geometrically this picture corresponds to the special choice of the
moving reference frame
$\{|N>,|T_1>,...,|T_{N-1}>,|T_1*>,...,|T_{N-1}*>\}$ on $CP(N-1)$
that only ``longitudinal'' component along $|N>$ is alive.

The Berry's formula (1.24) \cite{Berry198} being applied to the action state vector in the local coordinates $\pi^i$, i.e (\ref{4}), (\ref{5}) gives following equation for antisymmetric second-rank tensor
\begin{eqnarray}
V_{ik*}(\pi^i)= \Im \sum_{a=0}^{N-1} \{\frac{\partial
g^{a*}}{\partial \pi^i} \frac{\partial g^a}{\partial \pi^{k*}} -
\frac{\partial g^{a*}}{\partial \pi^{k*}} \frac{\partial
g^a}{\partial \pi^i} \} =\Im \sum_{a=0}^{N-1}\{T^a_i T^{a*}_k -
(T^a_k)^* T^{a}_i \} \cr = -\Im [(1+ \sum |\pi^s|^2) \delta_{ik}-
\pi^{i^*} \pi^k](1+ \sum |\pi^s|^2)^{-2}= - \Im G_{ik^*}
\label{form}.
\end{eqnarray}
It is simply the imaginary part of the Fubini-Study quantum metric
tensor. There are two important differences between
original Berry's formula referring to arbitrary parameters and this
2-form in local coordinates inherently related to eigen-problem.

1. The $V_{ik*}(\pi^i)$ is the singular-free expression.

2. It does not contain two eigen-values, say, $E_n, E_m$ explicitly,
but implicitly $V_{ik*}$ depends locally on the choice of single
$\lambda_p$ through the dependence in local coordinates
$\pi^i_{j(p)}$. In some sense it looks like ``degeneration" but now
the reason of the anholonomy larks in the curvature of $CP(N-1)$ and therefore
it has invariant character.

I formulate the problem to
find field equations for the $SU(N)$ parameters $\Omega^{\alpha}$
leading to the affine parallel transport of the Hamiltonian field
$H^i=\hbar \Omega^{\alpha}\Phi^i_{\alpha}$ where
\begin{equation}\label{16}
\Phi_{\sigma}^i = \lim_{\epsilon \to 0} \epsilon^{-1}
\biggl\{\frac{[\exp(i\epsilon \lambda_{\sigma})]_m^i g^m}{[\exp(i
\epsilon \lambda_{\sigma})]_m^j g^m }-\frac{g^i}{g^j} \biggr\}=
\lim_{\epsilon \to 0} \epsilon^{-1} \{ \pi^i(\epsilon
\lambda_{\sigma}) -\pi^i \}
\end{equation}
\cite{Le2}.
These field equations of motion for quantum system whose `particles' do not exist a priory
but they are becoming during the evolution. But first of all we
should to introduce the notion of the ``dynamical space-time" which
is arises due to the natural evolution or the objective measurement
of some dynamical variable.

The $CP(N-1)$ points serves as discriminators of physically distinguishable
quantum GCS. Let me assume that GCS described by the
local coordinates $(\pi^1,...,\pi^{N-1})$ and the coordinates $(\pi^1+\delta
\pi^1,...,\pi^{N-1}+\delta \pi^{N-1})$ correspond to the GCS
displaced due to measurement or evalution.

Local coordinates of the GCS gives the a firm geometric tool for
the description of quantum dynamics during interaction which used
for a measuring process or evolution. The question that I would like
to raise is as follows: {\it what ``classical field'', i.e. field in
space-time, corresponds to the transition from the original to the
displaced GCS?} In other words I would like to find the measurable
physical manifestation of the GCS in the form of ``field
shell", its space-time shape and its dynamics. The GCS dynamics
will be represented by (energy) frequencies distribution that are
not a priori given, but are defined by some field equations which
should established by means of variation problem applied to
operators represented by tangent vectors to $CP(N-1)$.

In order to build the qubit spinor $\eta$ of the quantum question
$\hat{Q}$ \cite{Le5} in the local basis
$\{|N>,|T_1>,...,|T_{N-1}>,|T_1*>,...,|T_{N-1}*>\}$ for the
measurement of the Hamiltonian $\hat{H}$ at corresponding GCS I
will use following equations
\begin{eqnarray}
\eta=\left(
  \begin{array}{cc}
    \alpha_{(\pi^1,...,\pi^{N-1})}  \\
    \beta_{(\pi^1,...,\pi^{N-1})} \\
  \end{array}
\right) = \left(
  \begin{array}{cc}
    \frac{<N|\hat{H}|\Psi>}{<N|N>}  \\
    \frac{<\Psi|\hat{H}|\Psi>}{<\Psi|\Psi>} \\
  \end{array}
\right)
\end{eqnarray}
Then from the infinitesimally close GCS
$(\pi^1+\delta^1,...,\pi^{N-1}+\delta^{N-1})$, whose shift is
induced by the interaction used for a measurement, one get a close
spinor $\eta+\delta \eta$ with the components
\begin{eqnarray}\label{514}
\eta + \delta \eta =\left(
  \begin{array}{cc}
    \alpha_{(\pi^1+\delta^1,...,\pi^{N-1}+\delta^{N-1})}  \\
    \beta_{(\pi^1+\delta^1,...,\pi^{N-1}+\delta^{N-1})} \\
  \end{array}
\right) = \left(
  \begin{array}{cc}
    \frac{<N|\hat{H'}|\Psi>}{<N|N>}  \\
    \frac{<\Psi|\hat{H'}|\Psi>}{<\Psi|\Psi>} \\
  \end{array}
\right)\end{eqnarray}
Here $\hat{H}=\hbar \Omega^{\alpha}\hat{\lambda}_{\alpha}$ is the lift of Hamiltonian vector field $H^i=\hbar \Omega^{\alpha} \Phi^i_{\alpha}$ from $(\pi^1,...,\pi^{N-1})$ and $\hat{H'}=\hbar(\Omega^{\alpha}+\delta \Omega^{\alpha}) \hat{\lambda}_{\alpha}$ is the lift of the parallel transported Hamiltonian vector field $H^i=\hbar \Omega^{\alpha} \Phi^i_{\alpha}$ (see below) from the infinitesimally close point
$(\pi^1+\delta^1,...,\pi^{N-1}+\delta^{N-1})$ back to the
$(\pi^1,...,\pi^{N-1})$ into the adjoint representation space.

Now one should find how the affine parallel transport connected
with the variation of coefficients $\Omega^{\alpha}$ in the
dynamical space-time associated with quantum question $\hat{Q}$.

The covariance relative transition from one GCS to another
\begin{eqnarray}
(\pi^1_{j(p)},...,\pi^{N-1}_{j(p)}) \rightarrow
(\pi^1_{j'(q)},...,\pi^{N-1}_{j'(q)})
\end{eqnarray}
and the covariant differentiation (relative Fubini-Study metric) of
vector fields provides the objective character of the ``quantum
question" $\hat{Q}$ and, hence, the quantum measurement. This serves
as a base for the construction of the dynamical space-time as it
will be shown below.

These two infinitesimally close spinors $\eta$ and $\eta+\delta
\eta$ may be connected with infinitesimal ``Lorentz spin transformations
matrix'' \cite{G}
\begin{eqnarray}
L=\left( \begin {array}{cc} 1-\frac{i}{2}\tau ( \omega_3+ia_3 )
&-\frac{i}{2}\tau ( \omega_1+ia_1 -i ( \omega_2+ia_2)) \cr
-\frac{i}{2}\tau
 ( \omega_1+ia_1+i ( \omega_2+ia_2))
 &1-\frac{i}{2}\tau( -\omega_3-ia_3)
\end {array} \right).
\end{eqnarray}
Then accelerations $a_1,a_2,a_3$ and angle velocities $\omega_1,
\omega_2, \omega_3$ may be found in the linear approximation from
the equation
\begin{eqnarray}\label{21}
\eta+\delta \eta = L \eta
\end{eqnarray}
as functions of the qubit spinor components of the quantum question
depending on local coordinates $(\pi^1,...,\pi^{N-1})$ involved in the $\delta \Omega^{\alpha}$ throughout field equations (\ref{25}). 

Hence the infinitesimal Lorentz transformations define small
``space-time'' coordinates variations. It is convenient to take
Lorentz transformations in the following form $ct'=ct+(\vec{x}
\vec{a}) d\tau, \quad \vec{x'}=\vec{x}+ct\vec{a} d\tau
+(\vec{\omega} \times \vec{x}) d\tau$, where I put
$\vec{a}=(a_1/c,a_2/c,a_3/c), \quad
\vec{\omega}=(\omega_1,\omega_2,\omega_3)$ \cite{G} in order to have
for $\tau$ the physical dimension of time. The expression for the
``4-velocity" $ V^{\mu}$ is as follows
\begin{equation}
V^{\mu}=\frac{\delta x^{\mu}}{\delta \tau} = (\vec{x} \vec{a},
ct\vec{a}  +\vec{\omega} \times \vec{x}) .
\end{equation}
The coordinates $x^\mu$ of points in dynamical space-time serve in
fact merely for the parametrization of deformations of the ``field
shell'' arising under its motion according to non-linear field
equations \cite{Le1,Le3}.

\section{Field equations in the dynamical space-time}
The energetic packet  - ``lump'' associated with the ``field
shell'' is now described locally by the Hamiltonian vector field
$\vec{H}=\hbar
\Omega^{\alpha}\Phi^i_{\alpha}\frac{\partial}{\partial \pi^i} + c.c
$. Our aim is to find the wave equations for $\Omega^{\alpha}$ in
the dynamical space-time intrinsically connected with the objective
quantum measurement (evolution).

At each point $(\pi^1,...,\pi^{N-1})$ of the $CP(N-1)$ one has an
``expectation value'' of the $\vec{H}$ defined by a measuring
device. But a displaced GCS may by reached along one of the
continuum paths. Therefore the comparison of two vector fields and
their ``expectation values'' at neighboring points requires some
natural rule. The comparison makes sense only for the same
lump represented by its ``field shell'' along some path. For
this reason one should have an identification procedure. The affine
parallel transport in $CP(N-1)$ of vector fields is a natural and
the simplest rule for the comparison of corresponding ``field
shells''.

The dynamical space-time coordinates $x^{\mu}$ may be introduced as
the state-dependent quantities, transforming in accordance with the
functionally local Lorentz transformations $\delta x^{\mu} = V^{\nu}
\delta \tau $ depend on the transformations of local reference frame
in $CP(N-1)$ as it was described in the previous paragraph.

Let me discuss now the self-consistent problem
\begin{equation}
V^{\mu} \frac{\partial \Omega^{\alpha}}{\partial x^{\mu} } = -
(\Gamma^m_{mn} \Phi_{\beta}^n+\frac{\partial
\Phi_{\beta}^n}{\partial \pi^n}) \Omega^{\alpha}\Omega^{\beta},
\quad \frac{d\pi^k}{d\tau}= \Phi_{\beta}^k \Omega^{\beta}
\end{equation}
arising under the condition of the affine parallel transport
\begin{eqnarray}
\frac{\delta H^k}{\delta \tau} &= &\hbar \frac{\delta
(\Phi^k_{\alpha} \Omega^{\alpha})}{\delta \tau}=0
\end{eqnarray}
of the Hamiltonian vector field.

The simplest case of
$CP(1)$ dynamics assumes $1\leq \alpha,\beta \leq3,\quad i,k,n=1$. The quasi-linear field equations in the case of the spherical symmetry being split into the
real and imaginary parts take the form
\begin{eqnarray}\label{25}
\matrix{ (r/c)\omega_t+ct\omega_r=-2\omega \gamma F(u,v), \cr
(r/c)\gamma_t+ct\gamma_r=(\omega^2 - \gamma^2) F(u,v), \cr u_t=
U_1(u,v,\omega,\gamma), \cr v_t= U_2(u,v,\omega,\gamma), }
\label{self_sys}
\end{eqnarray}
where $F(u,v), U_1(u,v,\omega,\gamma), U_2(u,v,\omega,\gamma)$ well defined functions and
$\pi=u+iv$.
\section{Reality, EPR and Schr\"odinger's Cat paradoxes}
 Reality is very wide philosophical category; it is even wider than the notion of matter. The definition of such kind of category is very problematic. The all philosophical battles between materialism and idealism concern just the criterion of reality. Particulary, macroscopic physical reality has a long history too.
 Contradictable development of quantum physics evoked new attempts of subjective idealism or/and even agnosticism to take over materialism and objective character of the scientific description of ``reality". Such philosophy sharply contradicted to Einstein's point of view. He was sure that the objective description (independent from observer) of quantum phenomenon without any reference to the agnostic ``uncontrol perturbation" should be achieved. Pursing this aim, Einstein, Podolsky and Rosen were insisted to give the definition of ``element of physical reality" \cite{EPR}.

 In order to understand the character of difficulties discovered in EPR article it will be interesting to recall firstly the Einstein's point of view on ``reality" \cite{Einstein4}. Einstein, discussing reality of gravitation field, notes that distinguishing ``real" and ``non-real"  has no meaning. He proposed instead to distinguish proper values of physical system (invariants) and values depending on coordinate description.

Nevertheless, Einstein in EPR article avoided his own point of view and he together with co-authors came to definition of the ``element of physical reality". This definition leads to conclusion that whether ordinary quantum scheme is not complete or some space-like interaction (``spooky action") between spatially distinguished observers is ``real".
``No reasonable definition of reality could be expected to permit this" \cite{EPR}.
In fact locality-separability \cite{Fine} contradicts to link of eigenfunctions-eigenvalues prescribed by standard quantum approach denying necessity to know dynamics of quantum transitions from one stationary state to another. This is the source of the indeterminism and primitive projective postulate treating measuring process as instant process that leads to simultaneous ``knowledge" of quantum state of any (even remote) subsystem. In fact one has not knowledge but only ``believe". There are however a number of evidences of quantum long-range correlations, i.e. some ``outlandish reality" denied by EPR does exist! This paradox may be resolved if we return to Einstein's initial point of view. Namely, applying his approach to quantum physics one needs to distinguish functional (state-dependent) invariants and values depending on representation (choice of the functional basis). Probably the absence of this detail is one of the main reasons of Einstein's complaint about EPR text written by Podolsky \cite{Fine}.

Functional invariant reflects the objective character of quantum state and its symmetries.
In EPR example the correlations arise due to space-time symmetries and corresponding conservation laws. They have a statistical character since the dynamical nature of the entanglement is hidden under standard QM approach. Dynamical nature of entanglement related to conservation laws for LDV's in state space. Quantum measurement of LCD's being understood as set of answers on quantum questions ``yes/no" create local dynamical space-time structure \cite{Le1,Le2,Le3,Le4,Le5}. Thereby instead of EPR criterion of reality one
may search invariants of entangled state and quantities depending of its representation.
It is clear that spatially non-local description of the entangled (extended) state could not be relativistically invariant. Namely, detection of two parts of single extended system (two remote solitons and even single ``big" soliton) are relativistically non-invariant procedure depending on the choice of spatial reference frame. A long time this fact was the main argument against non-local quantum field theory. However there are no natural reasons for requirements of conservation of the \textbf{global} causal relations and space-time locality in self-interacting extended systems (like well known paradoxically formulated \textbf{general} question about priority of chicken or egg). \textbf{For such kind of quantum systems, the relativity should be accompanied by super-relativity to the choice of functional reference frame} \cite{Le1,Le2,Kryukov}.  Namely, broken Lorentz symmetry widely discussed now (see, say, \cite{Kostel}), should be locally restored with help the affine parallel transport  of the local Hamiltonian in the projective Hilbert state space that leads to extended soliton-like solutions \cite{Le1,Le3}.

The EPR article appeals to intuitively clear picture of good localized particles which after interaction may be treated as non-interacting and hence independent after short time. Einstein and his co-authors in 1935 did not have such non-local self-interacting objects as solitons. But now soliton solution of good defined quantum system with spatially extended quantum state may replace this intuitive picture at least in the framework of some model. This leads to new ``outlandish reality" in physics. I would like to show the difference between such non-local objects and EPR original entangled states.

Let me assume that two entangled particles are described by some two-soliton solution, say
of the SG equation as follows
\begin{eqnarray}
\Psi(x,t)=4 \arctan \frac{e^{\eta_1} + e^{\eta_2}}{1+e^{\eta_1} e^{\eta_2}} ,
\end{eqnarray}
where $\eta_r = \kappa_r x - \nu_r t + \eta_r^0$, and $r=1,2$. This solution describes configuration with energy concentrated in two different space-time points but it depends on single pear $(x,t)$ and may be formally rewritten as series in some functional basis
\begin{eqnarray}
\Psi(x,t)=\sum_{n=1}^{\infty} \psi^n u_n(x,t),
\end{eqnarray}
where $u_n(x,t)$ are eigen-functions of dynamical variable $A$.
This decomposition sharply differs from EPR bi-local wave-function
\begin{eqnarray}
\Psi(x_1,x_2)=\sum_{n=1}^{\infty} \psi_n(x_2) u_n(x_1).
\end{eqnarray}
If one uses different basis $v_n(x,t)$ consists of eigen-functions of dynamical variable $B$, one will have the decomposition
\begin{eqnarray}
\Psi(x,t)=\sum_{n=1}^{\infty} \phi^n v_n(x,t).
\end{eqnarray}
First, it does not mean, however, that $u_n(x,t)$ describes one of the ``hump" and that Fourier components ${\psi^n }$ describes the state of the second ``hump". Second, one could not, therefore, conclude that the alternative measurement of dynamical variable B having eigen-functions $v_n(x,t)$ leads to the conclusion that the second ``hump" will be in the state ${\phi^n }$. We have here the dynamical entanglement for self-interacting non-separable soliton system.  Therefore the EPR conclusion about instant dependence of ``reality" from remote measurement for such kind of entanglement is not applicable but global Lorentz invariance will be broken. Non-locality of such ``elementary" particles leading to breakdown of Lorentz symmetry may be compensated by affine gauge field associated with parallel transport of the local Hamiltonian. Phenomenologically it may appear as new states or particles resulting ``deformation" of the Hamiltonian during the parallel transport and the continuous ``measurement" provided by ``quantum boosts" and ``quantum rotations" of local Lorentz reference frame. In other words: in order to avoid the contradiction with causality, the local Lorentz reference frame should be adapted during ``scanning along lump". Such local Lorentz reference frame has been built above whose ``quantum boosts" and ``quantum rotations" are defined by formulas (\ref{21}) and (\ref{25}). 
The parallel transport of the local Hamiltonian provides the ``self-identity" of extended object, i.e. the affine gauge fields couple the soliton-like system (\ref{25}) discussed in \cite{Le1,Le3}.

Assuming
$\omega=\rho \cos \psi,  \gamma=\rho \sin \psi$ and then  $\omega^2+\gamma^2=\rho^2=constant$, the two first
PDE's may be rewritten as follows:
\begin{equation}\label{PDE}
\frac{r}{c}\psi_t+ct\psi_r=F(u,v) \rho \cos \psi.
\end{equation}
The one of the exact solutions of this quasi-linear PDE is
\begin{eqnarray}\label{ex_sol}
\psi_{exact}(t,r)= \arctan \frac{\exp(2c\rho F(u,v)
f(r^2-c^2t^2))(ct+r)^{2F(u,v)}-1}{\exp(2c\rho F(u,v)
f(r^2-c^2t^2))(ct+r)^{2F(u,v)}+1},
\end{eqnarray}
where $f(r^2-c^2t^2)$ is an arbitrary function of the interval.

In order to keep physical interpretation of these equations I will
find the stationary solution for (\ref{PDE}). Let me put $\xi=r-ct$. Then
one will get ordinary differential equation
\begin{equation}\label{ODE}
\frac{d\Psi(\xi)}{d \xi} = -F(u,v) \rho \frac{\cos \Psi(\xi)}{\xi}.
\end{equation}
Two solutions
\begin{equation}
\Psi(\xi) =arctan(\frac{\xi^{-2M} e^{-2CM}-1}{\xi^{-2M} e^{-2CM}+1},
\frac{2\xi^{-M} e^{-2CM}}{\xi^{-2M} e^{-2CM}-1} ),
\end{equation}
where $M=F(u,v) \rho$ are concentrated in the vicinity of the
light-cone looks like solitary waves. Here we have example of relativistic non-local solution arose due to restoration of the Lorentz symmetry.

This solution is localizable in some functional space. I would like to emphasize that locality is formally achievable in any appropriate functional state space \cite{Kryukov}, but we wish to have the state space capable to distinguish dynamically different states of non-local in space-time solutions. The projective Hilbert state space in local coordinates apparently distinguishes states with different amplitudes and/or phases.

\section{Summary}
Taking into account the main target of Einstein which is not perfectly realized in the EPR article \cite{Fine,EPR}, the uncontrollable  perturbation  during a quantum measurement has been replaced to the coset structure $G/H=SU(N)/S[U(1) \times U(N-1)]=CP(N-1)$ of the quantum state deformation by the ``quantum force" \cite{Le2,Le3}.

In the framework of this model I may conclude:

1. Quantum measurement consists  of two procedures: the comparison of the local dynamical variables with the help of affine parallel transport in $CP(N-1)$ and the qubit-encoding of the quantum measurement.

2. Space-time emergences due to the qubit-encoding of the quantum measurement of local dynamical variables in projective Hilbert space.

3. Global Lorentz symmetry is broken and local unitary transformations
of the affine parallel transport of the local Hamiltonian in state space should restore it.

4. The superposition principle is realized locally in tangent space $T_{\pi}CP(N-1)$ at the solution of variational problem.

5. Identification of coset transformations $SU(N)/S[U(1) \times U(N-1)]$ and $CP(N-1)$ manifold of GCS is dynamical, i.e. state dependent. This means that a concrete form of the Cartan decomposition is state dependent, functionally local. Therefore vector fields representing local dynamical variable has invariant, physically objective character.

6. Schr\"odinger's Cat \cite{Schr2} paradox may be treated as a ``compact version of the EPR argument for incompleteness" \cite{Fine}.
 Furthermore, the Cat plays the role of two-level system (logical spin 1/2 \cite{Le2}, or qubit) whose 2 states ``yes" and ``no" fix a result of ``measurement". The continuous description of quantum evolution in $CP(N-1)$ and the covariant differentiation of local Hamiltonian provide the instant ``collapse" by local projection onto tangent space $T_{\pi}CP(N-1)$.   Following procedure of encoding the result of the ``measurement" by the lift in the tangent fibre bundle leads to the local dynamical space-time structure provided by moving local Lorentz reference frame.

7. The geometric formulation of QM being taken not as embellishment but as serious reconstruction, paves the way to new physical interpretation resolving old paradoxes (EPR, Schr\"odinger's Cat), namely: standard QM is incomplete and non-local. It requires reformulation in accordance with super-relativity like the classical mechanics was reformulated in accordance with Lorentz invariance of Maxwell equations.


\vskip 0.2cm
\begin{thebibliography}{99}
\bibitem{Le1}
P. Leifer, Annales de la Fondation Louis de Broglie, {\bf 32}, (1) 25
(2007).
\bibitem{Penrose}
R. Penrose, {\it The Road to Reality}, Alfred A.Knopf, New-York,
(2005).
\bibitem{Bonder}
Y. Bonder, arXiv:0801.2919v1 [gr-qc].
\bibitem{Einstein1}
A. Einstein, Ann. Phys. {\bf 17}, 891 (1905).
\bibitem{Einstein2}
A. Einstein, Ann. Phys. {\bf 49}, 769 (1916).
\bibitem{W}
T.D. Newton and E.P. Wigner, Rev. Mod. Phys., {\bf 21}, 400 (1949).
\bibitem{Heg}
G.C. Hegerfeldt, Phys.Rev.D{\bf 10}, 3320 (1974).
\bibitem{A98}
Y. Aharonov, {\it et al}, Phys.Rev. A{\bf 57}, 4130 (1998).
\bibitem{Kryukov}
A.A. Kryukov, Found. Phys. {\bf 36}, 175 (2006).
\bibitem{Schr1}
E. Schr\"odinger, Ann. Physik, \textbf{79}, 361 (1927).
\bibitem{dB1}
L. de Broglie, {\it On the Theory of Quanta}, A translation of ``Rechercher sur la Th\'eorie des Quanta" (Ann. de Phys., $10^e$ s\'erie, t.III (Janvier-F\'evrier 1925). by A.F. Kracklauer, 2004.
\bibitem{Le2}
P. Leifer, Found. Phys. {\bf 27}, (2) 261 (1997).
\bibitem{Berry198}
M.V. Berry, {\it Quantum Adiabatic Anholonomy}, Lectures.
\bibitem{Le3}
P. Leifer, arXiv:gr-qc/0503083.
\bibitem{G}
C.W. Misner, K.S. Thorne, J.A. Wheeler,{\it Gravitation}, W.H.Freeman
and Company, San Francisco, 1973.
\bibitem{EPR}
A. Einstein, B. Podolsky and N. Rosen, Phys.Rev. {\bf 47}, 777 (1935).
\bibitem{Einstein4}
A.Einstein, Naturwiss., {\bf 6}, 697 (1918).
\bibitem{Fine}
A. Fine, ``The Einstein-Podolsky-Rosen Argument in Quantum Theory", http://plato.stanford.edu/entries/qt-epr.
\bibitem{Le4}
P. Leifer, Found.Phys.Lett., {\bf 18}, (2) 195 (2005).
\bibitem{Le5}
P. Leifer, JETP Letters, {\bf 80}, (5) 367 (2004).
\bibitem{Kostel}
V.A. Kosteleck\'y, arXiv:0802.0581v1 [gr-qc].
\bibitem{Schr2}
E. Schr\"odinger, Naturewissenschafen 23: pp.807-812; 823-828; 844-849 (1935).

\end{thebibliography}
\end{document}